\documentclass[twocolumn]{IEEEtran}

\input{mysymbol.sty}
\usepackage{graphicx,dsfont}
\usepackage{epsfig}
\usepackage{epstopdf}
\usepackage{amsmath}
\usepackage{bbm}
\usepackage{amssymb}
\usepackage{wrapfig}
\usepackage{times,color}
\usepackage{cite,footnote,xspace,syntonly,algorithm,algorithmic,bm}
\usepackage{algorithm,algorithmic}
\usepackage[caption=false,font=footnotesize]{subfig}
\usepackage{mdwmath}
\usepackage{mdwtab}
\usepackage{eqparbox}

\def \A {{\mathbf{A}}}
\def \B {{\mathbf{B}}}

\def \I {{\mathbf{I}}}
\def \Y {{\mathbf{Y}}}
\def \X {{\mathbf{X}}}
\def \E {{\mathbf{E}}}

\def \M {{\mathbf{M}}}

\def \G {{\mathbf{G}}}
\def \U {{\mathbf{U}}}
\def \V {{\mathbf{V}}}

\def \y {{\mathbf{y}}}
\def \x {{\mathbf{x}}}
\def \e {{\mathbf{e}}}
\def \a {{\mathbf{a}}}

\long\def\symbolfootnote[#1]#2{\begingroup
\def\thefootnote{\fnsymbol{footnote}}
\footnote[#1]{#2}\endgroup} \psfull

\begin{document}


\title{\huge Dynamic Structural Equation Models for\\ Social Network Topology Inference$^\dag$}

\author{{\it Brian Baingana, \textit{Student Member}, \textit{IEEE}, Gonzalo~Mateos, \textit{Member}, \textit{IEEE}, \\
and Georgios~B.~Giannakis, \textit{Fellow}, \textit{IEEE}$^\ast$}}

\markboth{IEEE JOURNAL OF SELECTED TOPICS IN SIGNAL PROCESSING (SUBMITTED)}
\maketitle \maketitle \symbolfootnote[0]{$\dag$ Work in this paper was supported by the NSF ECCS Grants 
No. 1202135 and No. 1343248, and the NSF AST Grant No. 1247885.
Parts of the paper will appear in the {\it Proc. of
the of 5th Intl. Workshop on Computational Advances in Multi-Sensor Adaptive Processing}, Saint Martin, December 15-18, 2013.} 
\symbolfootnote[0]{$\ast$ The authors are with the Dept.
of ECE and the Digital Technology Center, University of
Minnesota, 200 Union Street SE, Minneapolis, MN 55455. Tel/fax:
(612)626-7781/625-4583; Emails:
\texttt{\{baing011,mate0058,georgios\}@umn.edu}}



\thispagestyle{empty}\addtocounter{page}{-1}
\begin{abstract}
Many real-world processes evolve in cascades over complex networks, whose topologies 
are often unobservable and change over time. However, 
the so-termed adoption times when blogs mention popular news items, 
individuals in a community catch an infectious disease, or 
consumers adopt a trendy electronics product are typically known,
and are implicitly dependent on the underlying network. 
To infer the network topology, a \textit{dynamic} 
structural equation model is adopted to capture the 
relationship between observed adoption times and the unknown edge weights. 
Assuming a slowly time-varying topology and leveraging the 
sparse connectivity inherent to social networks, 
edge weights are estimated by minimizing a sparsity-regularized 
exponentially-weighted least-squares criterion. 
To this end, solvers with complementary strengths 
are developed by leveraging (pseudo) real-time 
sparsity-promoting proximal gradient iterations, 
 the improved convergence rate of accelerated variants,
or reduced computational complexity of stochastic gradient descent.
Numerical tests with both synthetic and real data demonstrate the effectiveness of the 
novel algorithms in unveiling sparse dynamically-evolving 
topologies, while accounting for external influences in the adoption times. 
Key events in the recent succession of political leadership in North Korea, explain 
connectivity changes observed in the associated network inferred from global cascades
of online media.
\end{abstract}


%


\section{Introduction}
\label{sec:introduction}
Networks arising in natural and man-made settings
provide the backbone for the propagation of \emph{contagions} such as 
the spread of popular news stories, the adoption of buying trends among consumers, 
and the  spread of infectious diseases~\cite{rogers_book,kleinberg_book}. For example, a terrorist attack
may be reported within minutes on mainstream news websites.
An information cascade emerges because these websites' readership typically includes bloggers 
who write about the attack as well, influencing their own 
readers in turn to do the same. Although the times when ``nodes" get infected are often observable, the underlying 
network topologies over which cascades propagate are typically unknown and dynamic. 
Knowledge of the topology plays a crucial role for several reasons 
e.g., when social media advertisers select a small set of initiators
so that an online campaign can go viral, or when
healthcare initiatives wish to infer hidden needle-sharing networks of 
injecting drug users. As a general principle, network structural information 
can be used to predict the behavior of complex systems~\cite{kolaczyk_book}, such as 
the evolution and spread of information pathways in online media underlying e.g., 
major social movements and uprisings due to political conflicts~\cite{gomez}.
 
Inference of networks using temporal traces of infection events has recently
become an active area of research. According to the taxonomy in~\cite[Ch. 7]{kolaczyk_book},  
this can be viewed as a problem involving inference of \textit{association} networks. 
Two other broad classes of
network topology identification problems entail (individual) link prediction, or, tomographic
inference. Several prior approaches postulate
probabilistic models and rely on maximum likelihood estimation (MLE) to infer edge weights as
pairwise transmission rates between nodes \cite{gomez2},~\cite{meyers}. However,
these methods assume that the network does not change over time. A dynamic 
algorithm has been recently proposed to infer 
time-varying diffusion networks by solving an MLE problem via stochastic gradient descent iterations~\cite{gomez}. 
Although successful experiments on large-scale web data reliably uncover information
pathways, the estimator in~\cite{gomez} does not explicitly 
account for edge sparsity prevalent in social and information networks. 
Moreover, most prior approaches only attribute
node infection events to the network topology, and do not account for the influence of external 
sources such as a ground crew for a mainstream media website.

The propagation of a contagion is tantamount to \textit{causal} effects or interactions being excerted among 
entities such as news portals and blogs, consumers, or people susceptible to being infected with a contagious disease. 
Acknowledging this viewpoint, \textit{structural equation models} (SEMs) provide a 
general statistical modeling technique to estimate causal relationships among traits; see e.g.,~\cite{kaplan_book,pearl_book}. 
These directional effects are often 
not revealed by standard linear models that leverage symmetric associations between random
variables, such as those represented by covariances or correlations,~\cite{meinshausen},~\cite{friedman},~\cite{kolar},~\cite{daniele}. 
SEMs are attractive because of their 
simplicity and ability to capture edge directionalities. They have been widely adopted in 
many fields, such as economics, psychometrics~\cite{muthen}, social sciences~\cite{goldberger}, and genetics~\cite{liu,juan1}. 
In particular, SEMs have recently been proposed for \textit{static} gene regulatory network inference 
from gene expression data; see e.g., \cite{juan1,logsdon} and references therein.
However, SEMs have not been utilized to track the dynamics of causal effects among interacting nodes,
or, to infer the topology of time-varying directed networks. 

In this context, the present paper proposes a \textit{dynamic} SEM to account 
for directed networks over which
contagions propagate, and describes how node infection times depend on both
topological (causal) and external influences. Topological influences are
modeled in Section \ref{sec:model} as linear combinations of infection times
of other nodes in the network, whose weights correspond to entries in the 
time-varying asymmetric adjacency matrix. Accounting for external influences is
well motivated by drawing upon examples from online media, where established
news websites depend more on on-site reporting than blog references.  
External influence data is also useful for model identifiability, since it
has been shown necessary to resolve directional ambiguities~\cite{juan3}.
Supposing the network varies slowly with time, 
parameters in the proposed dynamic SEM are estimated adaptively
by minimizing a sparsity-promoting exponentially-weighted least-squares (LS) criterion (Section \ref{ssec:rls}).  
To account for the inherently sparse connectivity of social networks, an $\ell_1$-norm regularization 
term that promotes sparsity on the entries of the network adjacency matrix is incorporated in the cost function; see also
\cite{chen,kostas1,juan2,daniele}. 

A novel algorithm to jointly track
the network's adjacency matrix and the weights capturing the level of external influences is developed
in Section \ref{ssec:ista}, which
minimizes the resulting non-differentiable cost function via a proximal-gradient (PG)
solver; see e.g.,~\cite{boyd2,daubechies,beck}. The resulting
dynamic iterative shrinkage-thresholding algorithm (ISTA) 
is provably convergent, and offers parallel, closed-form, 
and sparsity-promoting updates per iteration. 
Proximal-splitting algorithms such as ISTA have been successfully 
adopted for various signal processing tasks~\cite{bauschke}, 
and for parallel optimization~\cite{combettes}. 
Further algorithmic improvements are outlined in Section \ref{sec:alg_improv}. 
These include enhancing the algorithms' rate of convergence 
through Nesterov's acceleration techniques~\cite{beck,nesterov1,nesterov2} (Section \ref{ssec:fista}),
and also adapting it for real-time operation (Section \ref{ssec:inexact_fista}).
When minimal computational complexity is at a premium, 
a stochastic gradient descent (SGD) algorithm is developed in Section \ref{ssec:stochgrad}, 
which adaptively minimizes an instantaneous (noisy) approximation
of the ensemble LS cost. Throughout,
insightful and useful extensions to the proposed algorithms that are not fully
developed due to space limitations are highlighted as remarks.

Numerical tests on synthetic network data demonstrate
the superior error performance of the developed algorithms,
and highlight their merits when compared to the 
sparsity-agnostic approach in~\cite{gomez} (Section \ref{ssec:synthetic}).  
Experiments in Section \ref{ssec:real} involve real temporal traces of popular
global events that propagated on news websites and blogs
in 2011~\cite{data}. Interestingly, topologies inferred from cascades associated to the
meme ``Kim Jong-un" exhibit an abrupt increase in the number of edges
following the appointment of the new North Korean ruler.

\noindent\textit{Notation}. Bold uppercase (lowercase) letters will denote matrices
(column vectors), while operators $(\cdot)^{\top}$, $\lambda_{\max}(\cdot)$, and
$\textrm{diag}(\cdot)$ will stand for matrix transposition, maximum eigenvalue, 
and diagonal matrix, respectively. The $N
\times N$ identity matrix will be represented by $\I_N$, while $\mathbf{0}_{N}$ will denote the $N
\times 1$ vector of all zeros, and $\mathbf{0}_{N \times
P}:=\mathbf{0}_{N} \mathbf{0}^\top_{P}$. The $\ell_p$ and Frobenius norms will be denoted by $\|\cdot\|_p$,
and $\|\cdot\|_F$, respectively.


\section{Network Model and Problem Statement}
\label{sec:model}
\begin{figure}[t]
\centering
\includegraphics[scale=0.4]{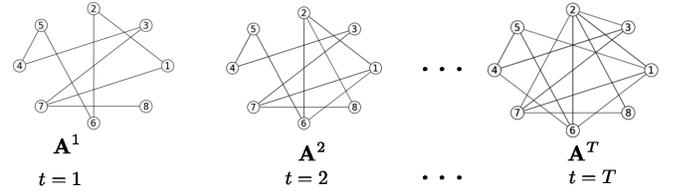}
\caption{Dynamic network observed across several time intervals. Note 
that few edges are added/removed in the transition from $t=1$ to $t=2$ 
(slowly time-varying network), and edges are depicted as undirected here for convenience.}
\label{fig:sgraphs}
\end{figure}

Consider a dynamic network with $N$ nodes observed
over time intervals $t = 1, \dots, T$, whose abstraction is a graph with topology described by an 
unknown, time-varying,  and weighted adjacency matrix $\A^t \in \mathbb{R}^{N \times N}$.
Entry $(i,j)$ of $\A^t$ (henceforth denoted by $a_{ij}^t$) is nonzero only if a directed edge connects
nodes $i$ and $j$ (pointing from $j$ to $i$) during the time interval $t$, as
illustrated in the $8$-node network in Fig. \ref{fig:sgraphs}. As a result, one in general
has $a_{ij}^t\neq a_{ji}^t$, i.e., matrix $\A^t$ is generally non-symmetric, which is suitable to model directed networks. 
For instance, if $i$ denotes a news blog maintained
by a journalism student, whereas $j$ represents the web portal of a mainstream newspaper, then it is likely
that $a_{ij}^t\gg a_{ji}^t\approx 0$ for those $t$ where $a_{ij}^t\neq 0$. Probably, the aforementioned 
directionality would have been reversed during Nov.-Dec. 2010, 
if $i$ instead represents the Wikileaks blog. Note that the model tacitly assumes that the network topology remains fixed
during any given time interval $t$, but can change across time intervals.

Suppose $C$ contagions propagate
over the network, and the difference between infection time of
node $i$ by contagion $c$ and the earliest observation time
is denoted by $y_{ic}^t$. In online media, $y_{ic}^t$
can be obtained by recording the time when website
$i$ mentions news item $c$.
For uninfected nodes at slot $t$, $y_{ic}^t$ is set to an arbitrarily large number.
Assume that the susceptibility $x_{ic}$ of node $i$ to
external (non-topological) infection by contagion $c$ is known and
time invariant over the observation interval. In the web context,  $x_{ic}$ can be set to the
search engine rank of website $i$ with respect to (w.r.t.) keywords 
associated with $c$. 

The infection time of node $i$ during interval $t$ is modeled according to the following
\emph{dynamic} structural equation model (SEM)
\begin{equation}
\label{eq:mps1}
y_{ic}^t = \sum\limits_{j \ne i} a_{ij}^t y_{jc}^t + b_{ii}^t x_{ic} + e_{ic}^t
\end{equation}
where  $b_{ii}^t$ captures
the time-varying level of influence of external sources, and $e_{ic}^t$
accounts for measurement errors and unmodeled dynamics. It follows from \eqref{eq:mps1}
that if $a_{ij}^t\neq 0$, then $y_{ic}^t $ is affected by the value of $y_{jc}^t$.
Rewriting \eqref{eq:mps1} for the entire network leads to the vector model
\begin{equation}
\label{eq:mps2}
\y_c^t = \A^t \y_c^t + \B^t \x_c + \e_c^t
\end{equation}
where the $N \times 1 $ vector $\y_c^t := \left[  y_{1c}^t, \dots, y_{Nc}^t \right]^{\top}$ 
collects the node infection times by contagion $c$ during interval $t$, and 
$\B^t := \text{diag}(b_{11}^t, \dots, b_{NN}^t)$. Similarly, 
$\x_c := \left[ x_{1c}, \dots, x_{Nc} \right]^{\top}$
and $\e_c^t := \left[ e_{1c}^t, \dots, e_{Nc}^t \right]^{\top}$. Collecting observations
for all $C$ contagions yields the dynamic matrix SEM
\begin{equation}
\label{eq:mps3}
\Y^t = \A^t \Y^t + \B^t \X + \E^t
\end{equation}
where $\Y^t := \left[  \y_1^t, \dots, \y_C^t \right]$, 
$\X := \left[ \x_1, \dots, \x_C \right]$, and
$\E^t := \left[  \e_1^t, \dots, \e_C^t\right] $
are all $N \times C$ matrices. Note that the same network topology $\A^t$
is adopted for all contagions, which is suitable e.g., when different
information cascades are formed around a common meme or trending
(news) topic in the Internet; see also the real data tests in Section \ref{ssec:real}.  

Given $\{ \Y^t \}_{t=1}^{T}$ and $\X$, the goal is to track
the underlying network topology $\{ \A^t \}_{t=1}^{T}$ and 
the effect of external influences $\{ \B^t \}_{t=1}^{T}$. 
To this end, the novel
algorithm developed in the next section assumes slow time variation
of the network topology and leverages the inherent sparsity
of edges that is typical of social networks.


\section{Topology Tracking Algorithm}
\label{sec:algorithms}
This section deals with a regularized LS approach to estimating
$\{\A^t,\B^t\}$ in \eqref{eq:mps3}. In a \emph{static} setting with all
measurements $\{ \Y^t \}_{t=1}^{T}$ available, one solves the batch problem
\begin{eqnarray}
\label{eq:ta1}
\nonumber
\{\hat{\A}, \hat{\B}\} = \underset{\A, \B}{\text{arg min}} & &  \frac{1}{2} \sum\limits_{t=1}^{T} \| \Y^t - \A \Y^t - \B \X  \|_F^2 + 
\lambda \| \A \|_1 \\
\text{s. to} & & a_{ii} = 0, \; b_{ij} = 0, \; \forall  i \ne j 
\end{eqnarray}
where
$\| \A \|_1 := \sum_{i,j} | a_{ij} |$ is a sparsity-promoting regularization,
and $\lambda > 0$ controls the sparsity level of $\hat{\A}$. Absence of 
a self-loop at node $i$ is enforced by the constraint
$a_{ii} = 0$, while having $b_{ij} = 0, \; \forall  i \ne j $, ensures that $\hat{\B}$
is diagonal as in \eqref{eq:mps2}. 

\begin{remark}[MLE versus LS]\label{rem:ml}
\normalfont If the errors $e_{ic}^t\sim\mathcal{N}(0,\sigma^2)$ in \eqref{eq:mps1} 
are modeled as independent and identically distributed (i.i.d.) Gaussian random variables, 
the sparsity-agnostic MLEs of the SEM parameters are obtained by solving
\begin{equation}
\label{eq:ml}
\min_{\A,\B} \sum\limits_{t=1}^{T} \left[\frac{1}{2} \| \Y^t - \A \Y^t - \B \X  \|_F^2 + 
C\sigma^2\log|\det(\I-\A)|\right]
\end{equation}
subject to the constraints in \eqref{eq:ta1}~\cite{juan1}. Different from regression linear models, LS is not maximum likelihood (ML) 
when it comes to Gaussian SEMs.
Sparsity can be accounted for in the ML formulation through $\ell_1$-norm regularization.
Here, the LS approach is adopted because of its universal applicability beyond Gaussian models,
and because MLE of SEM parameters gives rise to non-convex criteria [cf. \eqref{eq:ml}]. 
\end{remark}


\subsection{Exponentially-weighted LS estimator}
\label{ssec:rls}
In practice, measurements are typically acquired in a sequential manner and 
the sheer scale of social networks calls for estimation algorithms with
minimal storage requirements. Recursive solvers enabling
sequential inference of the underlying network topology are thus preferred. 
Moreover, introducing a ``forgetting factor" that assigns more 
weight to the most recent residuals makes it possible to track slow temporal variations of the
topology. Note that the batch estimator \eqref{eq:ta1} yields single estimates $\{\hat\A,\hat{\B}\}$ 
that best fit the data $\{\Y^t\}_{t=1}^T$ and $\X$ over the whole measurement horizon $t=1,\ldots, T$,
and as such \eqref{eq:ta1} neglects potential network variations across time intervals.

For $t = 1, \dots, T$, the sparsity-regularized exponentially-weighted 
LS estimator (EWLSE)
\begin{eqnarray}
\label{eq:ta2}
\nonumber
\{\hat{\A}^{t}, \hat{\B}^{t}\} = \underset{\A, \B}{\text{arg min}} &  &  \frac{1}{2} \sum\limits_{\tau=1}^{t} \beta^{t-\tau}   
\| \Y^\tau - \A \Y^\tau - \B \X  \|_F^2 \\ 
\nonumber 
& & \;\;\;  + \lambda_{t} \| \A \|_1 \\
\text{s. to} & &  a_{ii} = 0, \; b_{ij} = 0, \; \forall  i \ne j
\end{eqnarray}
where $\beta \in (0,1]$ is the forgetting factor that forms
estimates $\{\hat{\A}^{t}, \hat{\B}^{t}\}$ using all measurements acquired
until time $t$. Whenever $\beta<1$, past data are exponentially discarded thus enabling tracking of
dynamic network topologies. The first summand in the cost corresponds to an exponentially-weighted moving average 
(EWMA) of the squared model residuals norms. The EWMA can be seen as an average modulated 
by a sliding window of equivalent length $1/(1-\beta)$, which clearly grows as $\beta\to 1$. 
In the so-termed infinite-memory setting whereby $\beta=1$, \eqref{eq:ta2}
boils down to the batch estimator  \eqref{eq:ta1}. Notice that $\lambda_{t}$  is allowed
to vary with time in order to capture the generally changing edge sparsity level. In a linear regression context,
a related EWLSE was put forth in~\cite{juan2} for adaptive estimation of sparse signals; see also~\cite{kostas1}
for a projection-based adaptive algorithm. 

Before moving on to algorithms, a couple of remarks are in order.

\begin{remark}[Modeling slow network variations via sparsity]\label{rem:fusedlasso}
\normalfont To explicitly model slow topological variations across time intervals, a 
viable approach is to include an additional regularization term $\mu_t\|\A-\hat{\A}^{t-1}\|_1$
in the cost of \eqref{eq:ta2}. This way, the estimator penalizes deviations of the current topology
estimate relative to its immediate predecessor $\hat{\A}^{t-1}$. Through the tuning parameter $\mu_t$, 
one can adjust how smooth are the admissible topology variations from interval to interval.
With a similar goal but enforcing temporal smoothness via kernels with adjustable bandwidth, 
an  $\ell_1$-norm-regularized logistic regression approach
was put forth in~\cite{kolar}.
\end{remark}

\begin{remark}[Selection of $\lambda_t$]\label{rem:select_lambda}
\normalfont Selection of the (possibly time-varying) tuning parameter $\lambda_t$
is an important aspect of regularization methods such as \eqref{eq:ta2}, because
$\lambda_t$ controls the sparsity level of the inferred network and how its structure may change
over time. For sufficiently large values of $\lambda_t$ one obtains the trivial solution 
$\hat{\A}^t=\mathbf{O}_{N\times N}$, while increasingly more dense graphs are obtained
as $\lambda_t\to 0$. An increasing $\lambda_t$ will
be required for accurate estimation over extended time-horizons, since for 
$\beta\approx 1$ the norm of the LS term in \eqref{eq:ta2}
grows due to noise accumulation. This way the effect of the regularization term will be downweighted 
unless one increases $\lambda_t$ at a suitable rate, for instance proportional to $\sqrt{\sigma^2 t}$
as suggested by large deviation tail bounds when the errors are assumed $e_{ic}^t\sim\mathcal{N}(0,\sigma^2)$,
and the problem dimensions $N,C,T$ are sufficiently large~\cite{meinshausen,morteza,juan2}. 
In the topology tracking experiments of Section \ref{sec:experiments}, a time-invariant
value of $\lambda$ is adopted and typically chosen via trial and error to optimize the
performance. This is justified since smaller values of $\beta$ 
are selected for tracking network variations, which also implies that past data (and noise) are 
discarded faster, and the norm of the LS term in \eqref{eq:ta2}
remains almost invariant. As future research it would be interesting
to delve further into the choice of $\lambda_t$ using model selection
techniques such as cross-validation~\cite{juan1}, Bayesian information criterion (BIC) scores~\cite{kolar},
or the minimum description length (MDL) principle~\cite{nacho}, and investigate how this choice 
relates to statistical model consistency in a dynamic setting. 
\end{remark}


\subsection{Proximal gradient algorithm}
\label{ssec:ista}
Exploiting the problem structure in \eqref{eq:ta2}, a proximal gradient (PG) algorithm is developed
in this section to track the network topology;  see~\cite{boyd2}
for a comprehensive tutorial treatment on proximal methods. PG methods have been popularized
for $\ell_1$-norm regularized linear regression problems, through the class of iterative 
shrinkage-thresholding algorithms (ISTA); see e.g.,~\cite{daubechies,wright}. 
The main advantage of ISTA over off-the-shelf
interior point methods is its computational simplicity. Iterations boil down to matrix-vector multiplications
involving the regression matrix, followed by a soft-thresholding operation~\cite[p. 93]{hastie_book}.

In the sequel, an ISTA algorithm is developed for the sparsity regularized dynamic SEM 
formulation \eqref{eq:ta2} at time $t$. Based on this module, a (pseudo) real-time algorithm for tracking
the dynamically-evolving network topology over the horizon $t=1,\ldots, T$ is obtained as well. The resulting
algorithm's memory storage requirement and computational cost per data sample 
$\{\Y^t,\X\}$ does not grow with $t$. 

\noindent\textbf{Solving \eqref{eq:ta2} for a single time interval $t$.} 
Introducing the optimization variable $\V:=[\A\:\B]$, observe that the gradient of 
$f(\V):=\frac{1}{2} \sum_{\tau=1}^{t} \beta^{t-\tau}   \| \Y^\tau - \A \Y^\tau - \B \X  \|_F^2$
is Lipschitz continuous with a (minimum) Lipschitz constant
$L_f=\lambda_{\max}(\sum_{\tau=1}^t\beta^{t-\tau}[(\Y^\tau)^\top\:\:(\X)^\top]^\top[(\Y^\tau)^\top\:\:(\X)^\top])$, i.e.,
$\|\nabla f(\V_1)-\nabla f(\V_2)\|\leq L_f\|\V_1-\V_2\|$,
$\forall\:\V_1,\:\V_2$ in the domain of $f$. The Lipschitz constant is time varying, but
the dependency on $t$ is kept implicit for notational convenience. 
Instead of directly optimizing the cost in \eqref{eq:ta2}, PG 
algorithms minimize a sequence of overestimators evaluated at judiciously chosen points $\U$ (typically
the current iterate, or a linear combination of the two previous iterates as discussed in Section \ref{ssec:fista}).
From the Lipschitz continuity of $\nabla f$, for any $\V$ and $\U$ in the domain of $f$, it holds that
$f(\V)\leq Q_f(\U,\V):=f(\U)+\langle\nabla f(\U),\V-\U\rangle + (L_f/2)\|\V-\U\|_F^2$. 
Next, define $g(\V):=\lambda_t\|\A\|_{1} $ and form the quadratic approximation of the cost $f(\V)+g(\V)$
[cf. \eqref{eq:ta2}] at a given point $\U$
\begin{align}\label{eq:Q_approx}
Q(\V,\U):={}&{}Q_f(\V,\U)+g(\V)\nonumber\\
={}&{}\frac{L_f}{2}\|\V-\G(\U)\|_F^2+g(\V)\nonumber\\
&{}+f(\U)-\frac{\|\nabla f(\U)\|_F^2}{2L_f}
\end{align}
where $\G(\U):=\U-(1/L_f)\nabla f(\U)$, and clearly $f(\V)+g(\V)\leq Q(\V,\U)$ for any $\V$ and $\U$. 
Note that $\G(\U)$ corresponds to a gradient-descent
step taken from $\U$, with step-size equal to $1/L_f$. 

With $k=1,2,\ldots$ denoting
iterations, PG algorithms set $\U:=\V[k-1]$ and 
generate the following sequence of iterates
\begin{align}\label{eq:pg_iterates}
\V[k]{}:={}&\arg\min_\V Q(\V,\V[k-1])\nonumber\\
{}={}&\arg\min_\V
\left\{\frac{L_f}{2}\|\V-\G(\V[k-1])\|_F^2+g(\V)\right\}
\end{align}
where the second equality follows from the fact that the last two summands in 
\eqref{eq:Q_approx} do not depend on $\V$. The optimization problem \eqref{eq:pg_iterates} is known as the
\emph{proximal operator} of the function $g/L_f$ evaluated at $\G(\V[k-1])$, and is denoted
as $\textrm{prox}_{g/L_f}(\G(\V[k-1]))$. Henceforth adopting the notation $\G[k-1]:=\G(\V[k-1])$ for convenience, 
the PG iterations can be compactly rewritten as
\begin{align}\label{eq:pg_iterates_simple}
\V[k]=\textrm{prox}_{g/L_f}(\G[k-1]).
\end{align}
A key element to the success of PG algorithms stems from the possibility of 
efficiently solving the sequence of subproblems \eqref{eq:pg_iterates}, i.e., evaluating
the proximal operator. Specializing to \eqref{eq:ta2}, note that \eqref{eq:pg_iterates}
decomposes into
\begin{align}
\nonumber\A[k]{}:={}&\arg\min_\A
\left\{\frac{L_f}{2}\|\A-\G_A[k-1]\|_F^2+\lambda_t\|\A\|_1\right\}\\
\label{eq:pg_iterates_A}={}&\textrm{prox}_{\lambda_t\|\cdot\|_1/L_f}(\G_A[k-1])\\
\label{eq:pg_iterates_B}\B[k]{}:={}&\arg\min_\B
\left\{\|\B-\G_B[k-1]\|_F^2\right\}=\G_B[k-1]
\end{align}
subject to the constraints in \eqref{eq:ta2} which so far have been left implicit, and $\G:= [\G_A\: \G_B]$.
Because there is no regularization on the matrix $\B$, the corresponding update \eqref{eq:pg_iterates_B} boils-down to
a simple gradient-descent step. Letting $\mathcal{S}_{\mu}(\M)$ with $(i,j)$-th entry given
by $\textrm{sign}(m_{ij})\max(|m_{ij}|-\mu,0)$ denote the soft-thresholding operator, it follows
that $\textrm{prox}_{\lambda_t\|\cdot\|_1/L_f}(\cdot)=\mathcal{S}_{\lambda_t/L_f}(\cdot)$, 
e.g.,~\cite{daubechies,hastie_book}; so that
\begin{align}\label{eq:pg_iterates_soft}
\A[k]=\mathcal{S}_{\lambda_t/L_f}(\G_A[k-1]).
\end{align}

What remains now is to obtain expressions for the gradient of $f(\V)$ with respect
to $\A$ and $\B$, which are required to form the matrices $\G_A$ and $\G_B$. To this end, note
that by incorporating the constraints $a_{ii} = 0$ and $b_{ij} = 0, \; \forall  j \ne i$, $i=1,\ldots N,$ 
one can simplify the expression of $f(\V)$ as
\begin{align}\label{eq:f_rowwise}
f(\V):=\frac{1}{2} \sum_{\tau=1}^{t}\sum_{i=1}^N \beta^{t-\tau}   \| (\y_i^\tau)^\top - \a_{-i}^\top\Y_{-i}^\tau - b_{ii}\x_i^\top  \|_F^2
\end{align}
where $(\y_i^\tau)^\top$ and $\x_i^\top$ denote the $i$-th row of $\Y^\tau$ and $\X$, respectively;
while $\a_{-i}^\top$ denotes the $1\times (N-1)$ vector obtained by removing entry $i$
from the $i$-th row of $\A$, and likewise  $\Y_{-i}^\tau$ is the $(N-1)\times C$ matrix obtained
by removing row $i$ from $\Y^\tau$. It is apparent from \eqref{eq:f_rowwise} that $f(\V)$
is separable across the trimmed row vectors $\a_{-i}^\top$, and the scalar diagonal entries $b_{ii}$, $i=1,\ldots, N$.
The sought gradients are readily obtained as
\begin{align}
\nonumber\nabla_{\a_{-i}} f(\V)={}&-\sum_{\tau=1}^{t}\beta^{t-\tau}\Y_{-i}^\tau ( \y_i^\tau - (\Y_{-i}^\tau)^\top\a_{-i} - \x_i b_{ii} )\\
\nonumber\nabla_{b_{ii}} f(\V)={}&-\sum_{\tau=1}^{t}\beta^{t-\tau}( (\y_i^\tau)^\top - \a_{-i}^\top\Y_{-i}^\tau - b_{ii}\x_i^\top)\x_i.
\end{align}
At time interval $t$, consider the data-related EWMAs $\bm{\Sigma}^t:=\sum_{\tau=1}^{t}\beta^{t-\tau}\Y^\tau(\Y^\tau)^\top$, $\bm\sigma_i^t:=\sum_{\tau=1}^{t}\beta^{t-\tau}\Y^\tau\y_i^\tau$, and $\bar{\Y}^t:=\sum_{\tau=1}^{t}\beta^{t-\tau}\Y^\tau$. With
these definitions, the gradient expressions for $i=1,\ldots,N$ can be compactly expressed as
\begin{align}
\label{eq:nabla_f_ai}\nabla_{\a_{-i}} f(\V)={}&\bm{\Sigma}_{-i}^t\a_{-i}+\bar{\Y}^t_{-i}\x_i b_{ii}-\bm\sigma_{-i}^t\\
\label{eq:nabla_f_bii}\nabla_{b_{ii}} f(\V)={}&\a_{-i}^\top\bar{\Y}^t_{-i}\x_i+\frac{1-\beta^t}{1-\beta}b_{ii}\|\x_i\|_2^2-
(\bar{\y}^\tau_{i})^\top\x_i
\end{align}
where $(\bar{\y}^t_{i})^\top$ denotes the $i$-th row of $\bar{\Y}^t$, $\bar{\Y}^t_{-i}$ 
is the $(N-1)\times C$ matrix obtained by removing row $i$ from $\bar{\Y}^t$, and $\bm{\Sigma}_{-i}^t$
is the $(N-1)\times (N-1)$ matrix obtained by removing the $i$-th row and $i$-th column from $\bm{\Sigma}^t$.

From \eqref{eq:pg_iterates_B}-\eqref{eq:pg_iterates_soft} and 
\eqref{eq:nabla_f_ai}-\eqref{eq:nabla_f_bii}, the parallel ISTA iterations
\begin{align}
\label{eq:nabla_f_ai_eval}\nabla_{\a_{-i}} f[k]={}&\bm{\Sigma}_{-i}^t\a_{-i}[k]+\bar{\Y}^t_{-i}\x_i b_{ii}[k]-\bm\sigma_{-i}^t\\
\label{eq:nabla_f_bii_eval}\nabla_{b_{ii}}f[k]={}&\a_{-i}^\top[k]\bar{\Y}^t_{-i}\x_i+
\frac{(1-\beta^t)}{1-\beta}b_{ii}[k]\|\x_i\|_2^2-(\bar{\y}^t_{i})^\top\x_i\\
\label{eq:pg_iters_ai}\hspace{-0.5cm}\a_{-i}[k+1]={}&\mathcal{S}_{\lambda_t/L_f}\left(\a_{-i}[k]-(1/L_f)\nabla_{\a_{-i}} f[k]\right)\\
\label{eq:pg_iters_bii} b_{ii}[k+1]={}& b_{ii}[k]-(1/L_f)\nabla_{b_{ii}}f[k]
\end{align}
are provably convergent to the globally optimal solution $\{\hat{\A}^t, \hat{\B}^t\}$ 
of \eqref{eq:ta2}, as per the general convergence results available
for PG methods and ISTA in particular~\cite{daubechies,boyd2}.

Computation of the gradients in \eqref{eq:nabla_f_ai_eval}-\eqref{eq:nabla_f_bii_eval}
requires one matrix-vector mutiplication by $\bm{\Sigma}_{-i}^t$ and one by $\bar{\Y}^t_{-i}$,
in addition to three vector inner-products, plus a few (negligibly complex) scalar and vector
additions. Both the update of $b_{ii}[k+1]$ as well as the soft-thresholding operation
in \eqref{eq:pg_iters_ai} entail negligible computational complexity. All in all, the simplicity
of the resulting iterations should be apparent.
Per iteration, the actual rows of the adjacency matrix are obtained by 
zero-padding the updated $\a_{-i}[k]$, namely setting
\begin{equation}
\label{eq:zero_pad}
\a_i^\top[k]=[a_{-i,i1}[k]\ldots a_{-i,ii-1}[k]\:0\:a_{-i,ii}[k]\ldots a_{-i,iN}[k]].
\end{equation}
This way, the desired SEM parameter estimates at time $t$ are given
by $\hat{\A}^{t} = [\a_1^\top[k],\ldots,\a_N^\top[k]]^\top$ and $\hat{\B}^{t}=
\textrm{diag}(b_{11}[k],\ldots,b_{NN}[k])$, for $k$ large enough so that convergence
has been attained.

\begin{remark}[General sparsity-promoting regularization]\label{rem:general_sparse}
Beyond $g(\A)=\lambda_t\|\A\|_1$, the algorithmic framework here can 
accommodate more general \textit{structured sparsity}-promoting 
regularizers $\gamma(\A)$ as long as the resulting proximal operator $\textrm{prox}_{\gamma/L_f}(\cdot)$
is given in terms of scalar or (and) vector soft-thresholding operators.
In addition to the $\ell_1$-norm (Lasso penalty), this holds e.g., for the
sum of the $\ell_2$-norms of vectors with groups of non-overlapping entries of $\A$ (group Lasso penalty~\cite{yuan}),
or, a linear combination of the aforementioned two -- the so-termed hierarchical
Lasso penalty that encourages sparsity across and within the groups defined over $\A$~\cite{sprechmann}.
These types of regularization could be useful if one e.g., has a priori knowledge that
some clusters of nodes are more likely to be jointly (in)active~\cite{gomez}.
\end{remark}

\noindent\textbf{Solving \eqref{eq:ta2} over the entire time horizon $t=1,\ldots,T$.} 
To track the dynamically-evolving network topology, one can go ahead and solve
\eqref{eq:ta2} sequentially for each $t=1,\ldots,T$ as data arrive, 
using \eqref{eq:nabla_f_ai_eval}-\eqref{eq:pg_iters_bii}. (The procedure can also be adopted in a 
batch setting, when all $\{\Y^t\}_{t=1}^T$ are available in memory.) Because the network is assumed to vary slowly
across time intervals, it is convenient to warm-restart the ISTA iterations, that is, at time $t$ initialize 
$\{\A[0],\B[0]\}$ with the previous solution $\{\hat{\A}^{t-1}, \hat{\B}^{t-1}\}$. Since
the sought estimates are expected to be close to the initial points, one expects convergence to be 
attained after few iterations.  

To obtain the new SEM parameter estimates via \eqref{eq:nabla_f_ai_eval}-\eqref{eq:pg_iters_bii}, 
it suffices to update (possibly) $\lambda_t$ and the
Lipschitz constant $L_f$, as well as the data-dependent EWMAs $\bm{\Sigma}^t$ ($\bm{\sigma}_{i}^t$ is the $i$-th column
of $\bm{\Sigma}^t$), and $\bar{\Y}^t$. Interestingly, the potential growing-memory problem
in storing the entire history of data $\{\Y^t\}_{t=1}^T$  can be avoided by performing
the recursive updates
\begin{align}
\label{eq:ta7a}
\bm\Sigma^{t} ={}& \beta\bm\Sigma^{t - 1} + \Y^{t} ({\Y^{t}})^{\top} \\
\label{eq:ta7b}
\bar{\Y}^{t} ={}& \beta \bar{\Y}^{t - 1} + \Y^{t} .
\end{align}
Note that the complexity in evaluating the Gram matrix $\Y^{t} ({\Y^{t}})^{\top}$ dominates
the per-iteration computational cost of the algorithm.  To circumvent the need of
recomputing the Lipschitz constant per time interval (that in this case entails
finding the spectral radius of a data-dependent matrix), the step-size $1/L_f$ in 
\eqref{eq:pg_iters_ai}-\eqref{eq:pg_iters_bii} can be selected by a line search~\cite{boyd2}. One possible
choice is the backtracking step-size rule in~\cite{beck}, under which convergence of 
\eqref{eq:nabla_f_ai}-\eqref{eq:pg_iters_bii} to $\{\hat{\A}^t, \hat{\B}^t\}$ can be established as well.

\begin{algorithm}[t!]
    \caption{Pseudo real-time ISTA for topology tracking}
\label{alg1}
\begin{algorithmic}[1]
   \REQUIRE  $\left\lbrace \Y^{t} \right\rbrace_{t=1}^T$,  $\X$, $\beta$.
   \STATE Initialize $\hat{\A}^{0} = \mathbf{0}_{N\times N}, \hat{\B}^{0} = \bm\Sigma^{0} = \I_N,\bar{\Y}^{0}=\mathbf{0}_{N\times C}, \lambda_{0}$. 
   \FOR{$t = 1, \dots, T$}
   \STATE Update $\lambda_{t}$, $L_f$ and $\bm\Sigma^{t}$, $\bar{\Y}^{t}$ via \eqref{eq:ta7a}-\eqref{eq:ta7b}.
   \STATE Initialize $\A[0]=\hat{\A}^{t-1}$, $\B[0]=\hat{\B}^{t-1}$, and set $k=0$.
   \WHILE {not converged}
   \FOR {$i=1 \dots N$ (in parallel)} 
   \STATE Compute $\bm\Sigma_{-i}^{t}$ and $\bar{\Y}_{-i}^{t}$.
   \STATE Form gradients at $\a_{-i}[k]$ and $b_{ii}[k]$ via \eqref{eq:nabla_f_ai_eval}-\eqref{eq:nabla_f_bii_eval}.
   \STATE Update $\a_{-i}[k+1]$ via \eqref{eq:pg_iters_ai}.
   \STATE Update $b_{ii}[k+1]$ via \eqref{eq:pg_iters_bii}.   
   \STATE Update $\a_i[k+1]$ via \eqref{eq:zero_pad}.
   \ENDFOR
   \STATE $k = k + 1$.
   \ENDWHILE
   \RETURN $\hat{\A}^{t} = \A[k]$, $\hat{\B}^{t} = \B[k]$.
   \ENDFOR
\end{algorithmic}
\end{algorithm}

Algorithm \ref{alg1} summarizes the steps outlined in this section for tracking the dynamic 
network topology, given temporal traces of infection events $\{\Y^t\}_{t=1}^T$
and susceptibilities $\X$. It is termed \emph{pseudo real-time} ISTA, since in principle
one needs to run multiple (inner) ISTA iterations till convergence per time interval $t=1,\ldots, T$. 
This will in turn incur an associated delay, that may (or may not) be tolerable depending
on the specific network inference problem at hand. Nevertheless, numerical tests indicate that
in practice 5-10 inner iterations suffice for convergence; see also Fig. \ref{fig:jstsp_fig_fista_iterations} and 
the discussion in Section \ref{ssec:inexact_fista}. 

\begin{remark}[Comparison with the ADMM in~\cite{baingana}]\label{rem:ADMM}
\normalfont In a conference precursor to this paper~\cite{baingana}, an alternating-direction 
method of multipliers (ADMM) algorithm was put forth to estimate the dynamic SEM 
model parameters. While the basic global structure of the algorithm in~\cite{baingana}
is similar to Algorithm 1, ADMM is adopted (instead of ISTA) to solve \eqref{eq:ta2} per time $t=1,\ldots,T$.
To update $\a_{-i}[k+1]$, ADMM iterations require inverting the matrix $\bm\Sigma_{-i}^{t}+\I_{N-1}$,
that could be computationally demanding for very large networks. On the other hand, Algorithm 1 
is markedly simpler and more appealing for larger-scale problems.
\end{remark}


\section{Algorithmic Enhancements and Variants}
\label{sec:alg_improv}
This section deals with various improvements to Algorithm 1, that pertain to accelerating its 
rate of convergence and also adapting it for real-time operation in time-sensitive applications.
In addition, a stochastic-gradient algorithm useful when minimal computational complexity 
is at a premium is also outlined.


\subsection{Accelerated proximal gradient method and fast ISTA}
\label{ssec:fista}
In the context of sparsity-regularized inverse problems and general non-smooth optimization, 
there have been several recent efforts towards improving the sublinear global rate of convergence
exhibited by PG algorithms such as ISTA; see e.g.,~\cite{beck,nesterov1,nesterov2} and references therein. 
Since for large-scale problems first-order (gradient) methods are in many cases the 
only admissible alternative, the goal of these works has been to retain the computational simplicity
of ISTA while markedly enhancing its global rate of convergence.

Remarkable results in~\cite{nesterov2} assert that convergence speedups 
can be obtained through the so-termed \emph{accelerated} (A)PG algorithm. 
Going back to the derivations in the beginning of Section \ref{ssec:fista},  
APG algorithms generate the following sequence of iterates [cf. \eqref{eq:pg_iterates} and \eqref{eq:pg_iterates_simple}]
\begin{align} 
\V[k]=\arg\min_\V Q(\V,\U[k-1])=\textrm{prox}_{g/L_f}(\G(\U[k-1]))\nonumber
\end{align}
where
\begin{align}
\label{eq:linear_comb}\hspace{-0.2cm}\U[k]{}:={}&\V[k-1]+\left(\frac{c[k-1]-1}{c[k]}\right)(\V[k-1]-\V[k-2])\\
\label{eq:t_seq}c[k]{}={}&\frac{1+\sqrt{4c^2[k-1]+1}}{2}.
\end{align}
In words, instead of minimizing a quadratic approximation to the cost evaluated at $\V[k-1]$
as in ISTA [cf. \eqref{eq:pg_iterates}], the accelerated PG algorithm [a.k.a. fast (F)ISTA] utilizes a linear combination
of the previous two iterates $\{\V[k-1],\V[k-2]\}$. The iteration-dependent 
combination weights are function of the scalar sequence \eqref{eq:t_seq}. FISTA
offers quantifiable iteration complexity, namely a (worst-case) convergence rate guarantee of $\mathcal{O}(1/\sqrt{\epsilon})$ 
iterations to return an $\epsilon$-optimal solution measured by its objective value 
(ISTA instead offers $\mathcal{O}(1/\epsilon)$)~\cite{beck,nesterov2}.
Even for general (non-)smooth optimization, APG algorithms have been shown to be optimal
within the class of first-order (gradient) methods, in the sense that the aforementioned  worst-case convergence rate
cannot be improved~\cite{nesterov1,nesterov2}.

The FISTA solver for \eqref{eq:ta2} entails the following steps [cf. \eqref{eq:nabla_f_ai_eval}-\eqref{eq:pg_iters_bii}]
\begin{align}
\label{eq:tilde_ai_linear_comb}\tilde\a_{-i}[k]{}:={}&\a_{-i}[k]+
\left(\frac{c[k-1]-1}{c[k]}\right)(\a_{-i}[k]-\a_{-i}[k-1])\\
\label{eq:tilde_bii_linear_comb}\tilde b_{ii}[k]{}:={}& b_{ii}[k]+
\left(\frac{c[k-1]-1}{c[k]}\right)( b_{ii}[k]-b_{ii}[k-1])\\
\label{eq:nabla_f_tilde_ai_eval}\nabla_{\a_{-i}} f[k]={}&\bm{\Sigma}_{-i}^t\tilde\a_{-i}[k]+\bar{\Y}^t_{-i}\x_i \tilde b_{ii}[k]-\bm\sigma_{-i}^t\\
\label{eq:nabla_f_tilde_bii_eval}\nabla_{b_{ii}}f[k]={}&\tilde\a_{-i}^\top[k]\bar{\Y}^t_{-i}\x_i+
\frac{(1-\beta^t)}{1-\beta}\tilde b_{ii}[k]\|\x_i\|_2^2-(\bar{\y}^t_{i})^\top\x_i\\
\label{eq:apg_iters_ai}\hspace{-0.5cm}\a_{-i}[k+1]={}&\mathcal{S}_{\lambda_t/L_f}\left(\tilde \a_{-i}[k]-(1/L_f)\nabla_{\tilde\a_{-i}} f[k]\right)\\
\label{eq:apg_iters_bii} b_{ii}[k+1]={}& \tilde b_{ii}[k]-(1/L_f)\nabla_{\tilde b_{ii}}f[k]
\end{align}
where $c[k]$ is updated as in \eqref{eq:t_seq}. The overall (pseudo) real-time FISTA for tracking
the network topology is tabulated under Algorithm \ref{alg2}. As desired, the computational
complexity of Algorithms \ref{alg1} and \ref{alg2} is roughly the same. Relative to 
Algorithm \ref{alg1}, the memory requirements
are essentially doubled since one now has to store the two prior estimates of $\A$ and $\B$,
which are nevertheless sparse and diagonal matrices, respectively. Numerical tests in
Section \ref{sec:experiments} suggest that Algorithm \ref{alg2} exhibits the best performance when compared
to Algorithm \ref{alg1} and the ADMM solver of~\cite{baingana},
especially when modified to accommodate real-time processing requirements -- the subject dealt
with next.

\begin{algorithm}[t!]
    \caption{Pseudo real-time FISTA for topology tracking}
\label{alg2}
\begin{algorithmic}[1]
   \REQUIRE  $\left\lbrace \Y^{t} \right\rbrace_{t=1}^T$,  $\X$, $\beta$.
   \STATE Initialize $\hat{\A}^{0} = \mathbf{0}_{N\times N}, \hat{\B}^{0} = \bm\Sigma^{0} = 
   \I_N,\bar{\Y}^{0}=\mathbf{0}_{N\times C}, \lambda_{0}$. 
   \FOR{$t = 1, \dots, T$}
   \STATE Update $\lambda_{t}$, $L_f$ and $\bm\Sigma^{t}$, $\bar{\Y}^{t}$ via \eqref{eq:ta7a}-\eqref{eq:ta7b}.
   \STATE Initialize $\A[0]=\A[-1]=\hat{\A}^{t-1}, \B[0]=\B[-1]=\hat{\B}^{t-1}$, $c[0]=c[-1]=1$, and set $k=0$.
   \WHILE {not converged}
   \FOR {$i=1 \dots N$ (in parallel)} 
   \STATE Compute $\bm\Sigma_{-i}^{t}$ and $\bar{\Y}_{-i}^{t}$.
   \STATE Update $\tilde\a_{-i}[k]$ and $\tilde b_{ii}[k]$ via \eqref{eq:tilde_ai_linear_comb}-\eqref{eq:tilde_bii_linear_comb}.
   \STATE Form gradients at $\tilde\a_{-i}[k]$ and $\tilde b_{ii}[k]$ via \eqref{eq:nabla_f_tilde_ai_eval}-\eqref{eq:nabla_f_tilde_bii_eval}.
   \STATE Update $\a_{-i}[k+1]$ via \eqref{eq:apg_iters_ai}.
   \STATE Update $b_{ii}[k+1]$ via \eqref{eq:apg_iters_bii}.   
   \STATE Update $\a_i[k+1]$ via \eqref{eq:zero_pad}.
   \ENDFOR
   \STATE $k = k + 1$.
   \STATE Update $c[k]$ via \eqref{eq:t_seq}.
   \ENDWHILE
   \RETURN $\hat{\A}^{t} = \A[k]$, $\hat{\B}^{t} = \B[k]$.
   \ENDFOR
\end{algorithmic}
\end{algorithm}



\subsection{Inexact (F)ISTA for time-sensitive operation}\label{ssec:inexact_fista}

Additional challenges arise with real-time data collection, where analytics 
must often be performed ``on-the-fly'' as well as without an opportunity 
to revisit past entries. Online operation in delay-sensitive applications may not tolerate
running multiple inner (F)ISTA iterations per time interval, so that
convergence is attained for each $t$ as required by Algorithms \ref{alg1}
and \ref{alg2}. This section touches upon an interesting tradeoff that 
emerges with time-constrained data-intensive problems, 
where a high-quality answer that is obtained slowly can be less 
useful than a medium-quality answer that is obtained quickly. 
 
Consider for the sake of exposition a scenario where the underlying network processes are
stationary, or just piecewise stationary with sufficiently long coherence time for that matter. 
The rationale behind the proposed real-time algorithm hinges upon the fact that the solution of \eqref{eq:ta2} for each
$t=1,\ldots, T$ does not need to be super accurate in the aforementioned stationary setting, 
since it is just an intermediate step in the outer loop matched to the time-instants of data acquisition. 
This motivates stopping earlier the inner iteration which solves \eqref{eq:ta2} (cf. the \textbf{while} loop 
in Algorithms \ref{alg1} and \ref{alg2}), possibly even after a single soft-thresholding step, as detailed 
in the real-time Algorithm \ref{alg3}. Note that in this case the inner-iteration index $k$ coincides
with the time index $t$.  A similar adjustment can be made to the ISTA variant (Algorithm \ref{alg1}), 
and one can in general adopt a less aggressive approach by allowing a few (not just one) 
inner-iterations per $t$. 

A convergence proof of Algorithm \ref{alg3} in a stationary network setting will
not be provided here, and is left as a future research direction. Still, convergence will be 
demonstrated next with the aid of computer simulations. 
For the infinite-memory case [cf. $\beta=1$ in \eqref{eq:ta2}] 
and the simpler ISTA counterpart of Algorithm \ref{alg3} 
obtained when $c[t]=1,\:\forall t$, it appears possible to adapt the arguments in~\cite{mairal,morteza} 
to establish that the resulting iterations converge to a minimizer of the batch problem \eqref{eq:ta1}.
In the dynamic setting where the network is time-varying, then convergence is not expected to 
occur because of the continuous network fluctuations. Still, as with adaptive signal processing 
algorithms~\cite{solo_book} one would like to establish that the tracking error attains a bounded 
steady-state. These interesting and challenging problems are subject of
ongoing investigation and will be reported elsewhere.

\begin{algorithm}[t!]
    \caption{Real-time inexact FISTA for topology tracking}
\label{alg3}
\begin{algorithmic}[1]
   \REQUIRE  $\left\lbrace \Y^{t} \right\rbrace_{t=1}^T$,  $\X$, $\beta$.
   \STATE Initialize $\A[1] = \A[0] =\mathbf{0}_{N\times N}, \B[1] =\B[0] = \bm\Sigma^{0} = 
   \I_N,\bar{\Y}^{0}=\mathbf{0}_{N\times C}, c[1]=c[0]=1, \lambda_{0}$. 
   \FOR{$t = 1, \dots, T$}
   \STATE Update $\lambda_{t}$, $L_f$ and $\bm\Sigma^{t}$, $\bar{\Y}^{t}$ via \eqref{eq:ta7a}-\eqref{eq:ta7b}.
   \FOR {$i=1 \dots N$ (in parallel)} 
   \STATE Compute $\bm\Sigma_{-i}^{t}$ and $\bar{\Y}_{-i}^{t}$.
   \STATE Update $\tilde\a_{-i}[t]$ and $\tilde b_{ii}[t]$ via \eqref{eq:tilde_ai_linear_comb}-\eqref{eq:tilde_bii_linear_comb}.
   \STATE Form gradients at $\tilde\a_{-i}[t]$ and $\tilde b_{ii}[t]$ via \eqref{eq:nabla_f_tilde_ai_eval}-\eqref{eq:nabla_f_tilde_bii_eval}.
   \STATE Update $\a_{-i}[t+1]$ via \eqref{eq:apg_iters_ai}.
   \STATE Update $b_{ii}[t+1]$ via \eqref{eq:apg_iters_bii}.   
   \STATE Update $\a_i[t+1]$ via \eqref{eq:zero_pad}.
   \ENDFOR
   \STATE Update $c[t+1]$ via \eqref{eq:t_seq}.
   \RETURN $\hat{\A}^{t}=\A[t+1], \hat{\B}^{t}=\B[t+1]$.
   \ENDFOR
\end{algorithmic}
\end{algorithm}

\begin{figure}[t]
\centering
\includegraphics[scale=0.65]{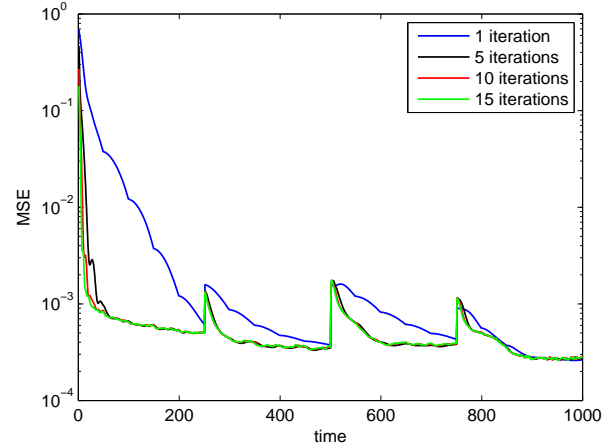}
\caption{MSE (i.e., $\sum_{i,j} (\hat{a}_{ij}^t - a_{ij}^t)^2/N^2 $) performance of Algorithm \ref{alg2} versus time. 
For each $t$, problem \eqref{eq:ta2} is solved ``inexactly'' for $k=1$ (Algorithm \ref{alg3}), $5$, $10$, and $15$ inner
iterations. It is apparent that $k=5$ iterations suffice to attain convergence to the minimizer of $\eqref{eq:ta2}$ per $t$, 
especially after a short transient where the warm-restarts offer increasingly better initializations.}
\label{fig:jstsp_fig_fista_iterations}
\end{figure}

For synthetically-generated data according to the setup described 
in Section \ref{ssec:synthetic}, Fig. \ref{fig:jstsp_fig_fista_iterations} shows the time
evolution of Algorithm 2's mean-square error (MSE) estimation performance. 
For each time interval $t$, \eqref{eq:ta2} is solved ``inexactly'' after 
running only $k=1,5,10$ and $15$ inner iterations. Note that the case $k=1$ 
corresponds to Algorithm \ref{alg3}. Certainly $k=10$ iterations suffice for the
FISTA algorithm to converge to the minimizer of \eqref{eq:ta2}; the curve for
$k=15$ is identical. Even with $k=5$ the obtained performance is
satisfactory for all practical purposes, especially
after a short transient where the warm-restarts offer increasingly better initializations.
While Algorithm \ref{alg3} shows a desirable convergent behavior, it seems that
this example's network coherence time of $t=250$ time intervals is too short
to be tracked effectively. Still, if the network changes are sufficiently smooth
as it occurs at $t=750$, then the real-time algorithm is able to estimate the network
reliably.

\subsection{Stochastic-gradient descent algorithm}
\label{ssec:stochgrad}
A stochastic gradient descent (SGD) algorithm is developed in this section,
which operates in real time  and can track the (slowly-varying) underlying 
network topology. Among all algorithms developed so far, the SGD iterations
incur the least computational cost. 

Towards obtaining the SGD algorithm, consider $\beta=0$ in \eqref{eq:ta2}.
The resulting cost function can be expressed as $f_t(\V)+g(\V)$, where $\V:=[\A\:\B]$ and 
$f_{t}(\V):=(1/2)\| \Y^t - \A \Y^t - \B \X  \|_F^2$ only accounts for the data acquired at time interval 
$t$. Motivated by computational simplicity, the ``inexact'' gradient descent
plus soft-thresholding ISTA iterations yield the following updates
\begin{align}
\label{eq:nabla_f_t_ai_eval}\nabla_{\a_{-i}} f_t[t]={}&\Y_{-i}^t \left((\Y_{-i}^t)^\top\a_{-i}[t] + \x_i b_{ii}[t] - \y_i^t\right)\\
\label{eq:nabla_f_t_bii_eval}\nabla_{b_{ii}}f_t[t]={}&  \a_{-i}^\top[t]\Y_{-i}^t\x_i + b_{ii}[t]\|\x_i\|^2-(\y_i^t)^\top\x_i\\
\label{eq:sgd_iters_ai}\hspace{-0.5cm}\a_{-i}[t+1]={}&\mathcal{S}_{\lambda_t/\eta}\left(\a_{-i}[t]-\eta\nabla_{\a_{-i}} f_t[t]\right)\\
\label{eq:sgd_iters_bii} b_{ii}[t+1]={}& b_{ii}[t]-\eta\nabla_{b_{ii}}f_t[t].
\end{align}
Compared to the parallel ISTA iterations in Algorithm \ref{alg1} [cf. \eqref{eq:nabla_f_ai_eval}-\eqref{eq:pg_iters_ai}],
three main differences are noteworthy: (i) iterations $k$ are merged with the time intervals $t$ of data acquisition;
(ii) the stochastic gradients $\nabla_{\a_{-i}} f_t[t]$ and $\nabla_{b_{ii}}f_t[t]$ involve the (noisy) data  
$\{\Y^{t} ({\Y^{t}})^{\top},\Y^{t}\}$ instead of their time-averaged counterparts $\{\bm\Sigma^{t},\bar{\Y}^{t}\}$;
and (iii) a generic constant step-size $\eta$ is utilized for the gradient descent steps.

\begin{algorithm}[t!]
    \caption{SGD algorithm for topology tracking}
\label{alg4}
\begin{algorithmic}[1]
   \REQUIRE  $\left\lbrace \Y^{t} \right\rbrace_{t=1}^T$,  $\X$, $\eta$.
   \STATE Initialize $\A[1] =\mathbf{0}_{N\times N}, \B[1]  = \I_N, \lambda_{1}$. 
   \FOR{$t = 1, \dots, T$}
   \STATE Update $\lambda_{t}$.
   \FOR {$i=1 \dots N$ (in parallel)}    
   \STATE Form gradients at $\a_{-i}[t]$ and $b_{ii}[t]$ via \eqref{eq:nabla_f_t_ai_eval}-\eqref{eq:nabla_f_t_bii_eval}.
   \STATE Update $\a_{-i}[t+1]$ via \eqref{eq:sgd_iters_ai}.
   \STATE Update $b_{ii}[t+1]$ via \eqref{eq:sgd_iters_bii}.   
   \STATE Update $\a_i[t+1]$ via \eqref{eq:zero_pad}.
   \ENDFOR
   \RETURN $\hat{\A}^{t}=\A[t+1], \hat{\B}^{t}=\B[t+1]$.
   \ENDFOR
\end{algorithmic}
\end{algorithm}


The overall SGD algorithm is tabulated under Algorithm \ref{alg4}. Forming the gradients 
in \eqref{eq:nabla_f_t_ai_eval}-\eqref{eq:nabla_f_t_bii_eval}
requires one matrix-vector mutiplication by $(\Y^t_{-i})^\top$ and two by $\Y^t_{-i}$. These 
multiplications dominate the per-iteration computational complexity of Algorithm \ref{alg4},
justifying its promised simplicity. Accelerated versions could be developed as well, at the
expense of marginal increase in computational complexity and doubling the memory requirements.

To gain further intuition on the SGD algorithm developed, consider the online learning
paradigm under which the network topology inference problem is to minimize the 
expected cost $E[ f_t(\V)+g(\V)]$ (subject to 
the usual constraints on $\V=[\A\:\B]$). The expectation is taken w.r.t. the \emph{unknown} 
probability distribution of the data. In lieu of the expectation, the 
approach taken throughout this paper is to minimize the empirical cost
$C^T(\V):=(1/T)[\sum_{t=1}^T f_t(\V)+g(\V)]$. Note that for $\beta=1$, the minimizers of $C^T(\V)$
coincide with \eqref{eq:ta1} since the scaling by $1/T$ does not affect the optimal
solution. For $\beta<1$, the cost $C_\beta^T(\V):=\sum_{t=1}^T\beta^{T-t} f_t(\V)+ g(\V)$
implements an EWMA which ``forgets'' past data and allows tracking.  In all cases, the rationale is that
by virtue of the law of large numbers, if data $\{\Y^t\}_{t=1}^T$ are stationary, solving
$\lim_{T\to\infty}\min_{\V}C^T(\V)$ yields the desired solution to the expected cost. 

A different approach to achieve this same goal -- typically with reduced computational complexity -- is to drop the
expectation (or the sample averaging operator for that matter), and update the estimates via a stochastic (sub)gradient iteration
$\V(t)=\V(t-1)-\eta \partial \{f_t(\V)+ g(\V)\}|_{\V=\V[t-1]}$. The subgradients with respect to $\a_{-i}$
are
\begin{align}\label{eq:partial_f_t_ai_eval}
\nonumber\partial_{\a_{-i}} f_t[t]={}&\Y_{-i}^t \left((\Y_{-i}^t)^\top\a_{-i}[t] + \x_i b_{ii}[t] - \y_i^t\right)\\
&+\lambda_t\textrm{sign}(\a_{-i}[t])
\end{align}
so the resulting algorithm has the drawback of (in general) not providing 
sparse solutions per iteration; see also~\cite{chen} for a 
sparse least-mean squares (LMS) algorithm. For that reason, the approach
here is to adopt the proximal gradient (ISTA) formalism to tackle the minimization of
the  instantaneous costs $f_t(\V)+ g(\V)$, and yield sparsity-inducing
soft-thresholded updates \eqref{eq:sgd_iters_ai}. Also acknowledging the limitation
of subgradient methods to yield sparse solutions,  related ``truncated gradient" updates 
were advocated for sparse online learning in~\cite{langford}.


\section{Numerical Tests}
\label{sec:experiments}

Performance of the proposed algorithms is
assessed in this section via computer simulations using both
synthetically-generated network data, and real traces of information cascades collected from
the web~\cite{data}.


\subsection{Synthetic data}
\label{ssec:synthetic}

\begin{figure}[t]
\centering
\includegraphics[scale=0.6]{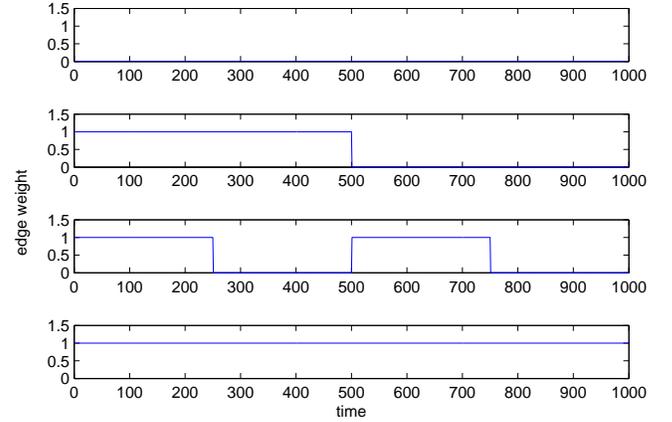}
\caption{Nonsmooth variation of synthetically-generated edge weights of the 
time-varying network. For each edge, one of the four depicted profiles is chosen uniformly at
random.}
\label{fig:weights}
\end{figure}

\begin{figure}[t]
\centering
\includegraphics[scale=0.5]{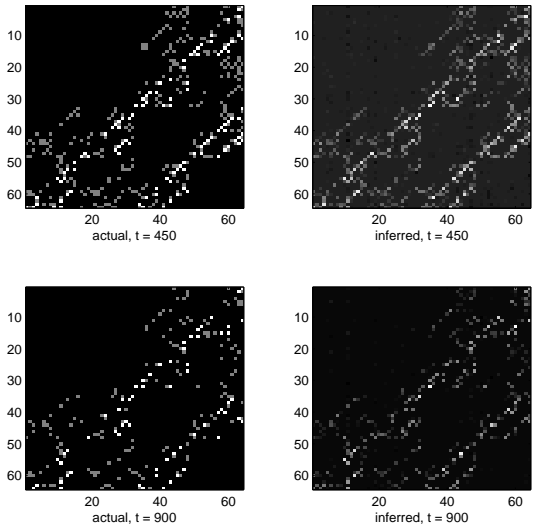}
\caption{Actual adjacency matrix $\hat{\A}^t$ and corresponding estimate $\hat{\A}^t$ obtained
using pseudo real-time ISTA (Algorithm \ref{alg1}), at time intervals $t=450$ and $t=900$.}
\label{fig:hmaps}
\end{figure}

\begin{figure}[t]
\centering
\includegraphics[scale=0.5]{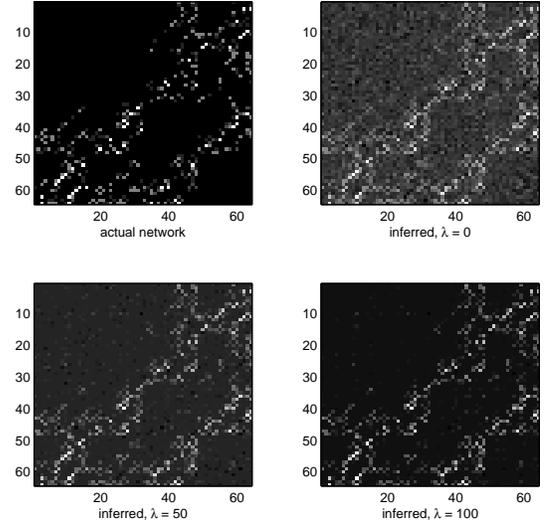}
\caption{Actual adjacency matrix at $t=900$ compared with the inferred
adjacency matrices using pseudo real-time FISTA (Algorithm \ref{alg2}), with $\lambda_t = \lambda$ for all
$t$ and $\lambda=0$, $\lambda=50$, and $\lambda=100$. While $\lambda=0$ and $\lambda=50$
markedly overestimate the support set associated with the true network edges,  the value $\lambda=100$ in this
case appears to be just about right.}
\label{fig:jstsp_fig_fista_hmaps_lambda}
\end{figure}

\noindent\textbf{Data generation.} Numerical tests on synthetic network data are conducted 
here to evaluate the tracking ability and compare Algorithms \ref{alg1}-\ref{alg4}. From a  ``seed graph" with adjacency matrix
\begin{equation*}
\mathbf{M} = \left(
\begin{array}{cccc}
0 & 0 & 1 & 1 \\
0 & 0 & 1 & 1 \\
0 & 1 & 0 & 1 \\
1 & 0 & 1 & 0
\end{array}
\right)
\end{equation*}
a Kronecker graph  of size $N=64$ nodes was generated as described in
\cite{kron}.\footnote{The Matlab implementation of
Algorithms \ref{alg1}-\ref{alg4} used here can handle networks of several
thousand nodes. Still a smaller network is analyzed since results
are still representative of the general behavior, and offers better visualization of the results in e.g.,
the adjacency matrices in Figs. \ref{fig:hmaps} and \ref{fig:jstsp_fig_fista_hmaps_lambda}.} The resulting
nonzero edge weights of $\A^t$ were allowed to vary over $T=1,000$ intervals 
under $3$ settings:
i) i.i.d. Bernoulli(0.5) random variables;
ii) random selection of the edge-evolution pattern uniformly from a set of four smooth functions: $a_{ij}(t) = 0.5+0.5\text{sin}(0.1t)$, 
$a_{ij}(t) = 0.5+0.5\text{cos}(0.1t)$,  $a_{ij}(t) = e^{-0.01t}$, and $a_{ij}(t) = 0$; and iii)
random selection of the edge-evolution pattern uniformly from a set of four nonsmooth functions shown in Fig. \ref{fig:weights}.

The number of contagions was set to $C=80$, and $\X$ 
was formed with i.i.d. entries uniformly distributed over
$[0,3]$.  Matrix $\B^t$ was set to $\text{diag}(\mathbf{b}^t)$, where 
$\mathbf{b}^t \in \mathbb{R}^N$ is a standard Gaussian random vector. During time interval $t$,
infection times were generated synthetically as $\Y^t = (\mathbf{I}_N - 
\A^t)^{-1} (\B^t\X + \E^t)$, where $\E^t$ is a standard Gaussian random matrix.

\begin{figure}[t]
\centering
\includegraphics[scale=0.6]{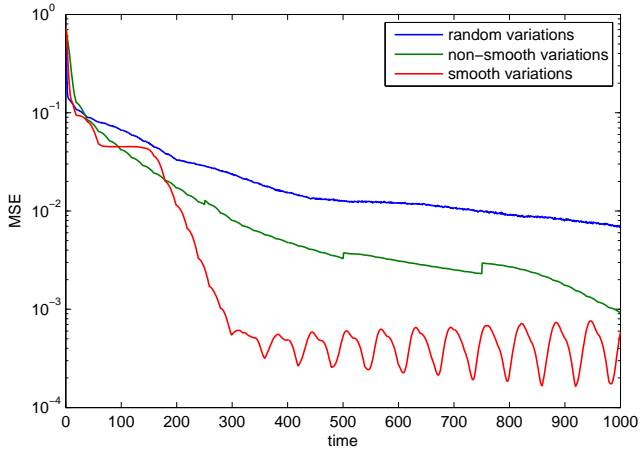}
\caption{MSE versus time obtained using pseudo real-time ISTA (Algorithm \ref{alg1}), 
for different edge evolution patterns.}
\label{fig:mse}
\end{figure}

\noindent\textbf{Performance evaluation.} With $\beta = 0.98$,
Algorithm \ref{alg1} was run after initializing the relevant variables as described
in the algorithm table (cf. Section \ref{ssec:ista}), 
and setting $\lambda_0 = 25$. In addition, $\lambda_{t} = \lambda_0$ for $t = 1, \dots, T$
as discussed in Remark \ref{rem:select_lambda}. 
Fig. \ref{fig:mse} shows the evolution of the mean-square error (MSE), 
$ \sum_{i,j} (\hat{a}_{ij}^t - a_{ij}^t)^2/N^2 $. As expected, the best 
performance was obtained when the temporal evolution of edges followed
smooth functions. Even though the Bernoulli evolution 
of edges resulted in the
highest MSE, Algorithm 1 still tracked the underlying topology 
with reasonable accuracy as depicted in 
the heat maps of the inferred adjacency matrices; see Fig. \ref{fig:hmaps}.

Selection of a number of parameters is critical
to the performance of the developed algorithms. In order
to evaluate the effect of each parameter on the network
estimates, several tests were conducted by tracking the non-smooth
network evolution using Algorithm \ref{alg2}
with varying parameter values. To illustrate the importance
of leveraging sparsity of the edge weights, 
Fig. \ref{fig:jstsp_fig_fista_hmaps_lambda} depicts heatmaps
of the adjacency matrices inferred at $t=900$, with $\lambda$ set to
$0, 50,$ and $100$ for all time intervals. Comparisons with
the actual adjacency matrix reveal that increasing $\lambda$ progressively
refines the network estimates by driving erroneously detected 
nonzero edge weights to $0$. Indeed, the value $\lambda=100$ in this
case appears to be just about right, while smaller values markedly overestimate
the support set associated with the edges present in the actual network.

\begin{figure}[t]
\centering
\includegraphics[scale=0.6]{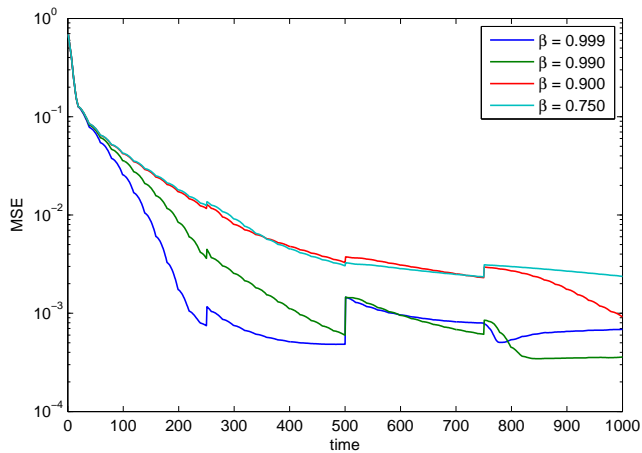}
\caption{MSE performance of the pseudo real-time FISTA (Algorithm \ref{alg2}) versus time, 
for different values of the forgetting factor $\beta$.}
\label{fig:jstsp_fig_fista_betas}
\end{figure}

\begin{figure}[t]
\centering
\includegraphics[scale=0.65]{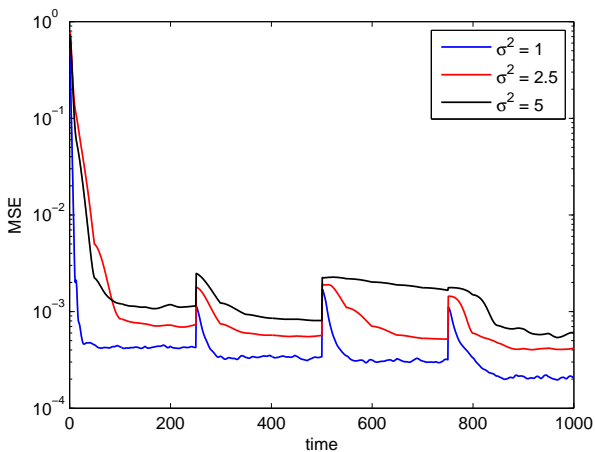}
\caption{MSE performance of the pseudo real-time FISTA (Algorithm \ref{alg2}) versus time, 
for different values of the noise variance $\sigma^2$.}
\label{fig:jstsp_fig_fista_noise}
\end{figure}

Fig. \ref{fig:jstsp_fig_fista_betas} compares 
the MSE performance of Algorithm \ref{alg2} for 
$\beta \in \{0.999, 0.990, 0.900, 0.750\}$. As expected, the
MSE associated with values of $\beta$ approaching $1$ degrades more
dramatically when changes occur within the network (at time intervals
$t=250$, $t=500$, and $t=750$ in this case; see Fig. \ref{fig:weights}). The MSE spikes observed when 
$\beta \in \{0.999, 0.990\}$ are a manifestation of the slower rate of adaptation
of the algorithm for these values of the forgetting factor. In this experiment,
$\beta = 0.990$ outperformed the rest for $t > 500$. In addition, comparisons 
of the MSE performance in the presence of increasing
noise variance are depicted in Figure \ref{fig:jstsp_fig_fista_noise}.
Although the MSE values are comparable during the initial stages
of the topology inference process, as expected higher noise levels lead to MSE performance
degradation in the long run.

\begin{figure}[t]
\centering
\includegraphics[scale=0.6]{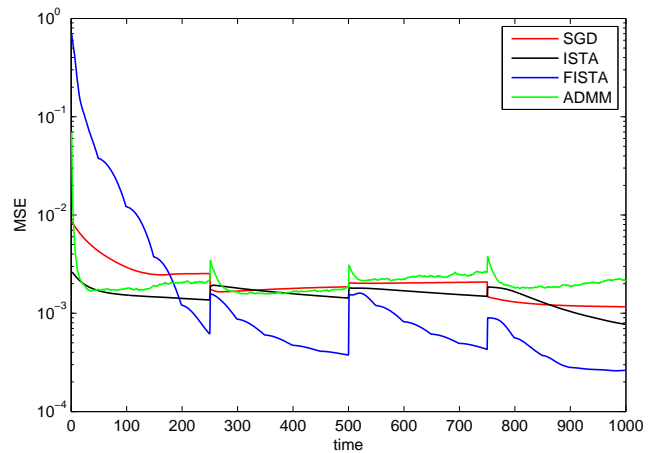}
\caption{MSE performance of the real-time algorithms versus time. Algorithms \ref{alg3} (real-time FISTA)
and \ref{alg4} (SGD), as well as inexact versions of Algorithm \ref{alg1} (ISTA) and the ADMM solver
in~\cite{baingana} are compared.}
\label{fig:jstsp_fig_sg_ista_fista_admm_online}
\end{figure}

Finally, a comparison of the real-time version of the different algorithms 
was carried out when tracking 
the synthetic time-varying network with non-smooth
edge variations.
Specifically, the real-time (inexact) counterparts of ISTA, FISTA (cf. Algorithm \ref{alg3}), SGD (cf.
Algorithm \ref{alg4}), and a suitably modified version of the
ADMM algorithm developed in \cite{baingana} were run
as suggested in Section \ref{ssec:inexact_fista}, i.e., eliminating the inner 
\textbf{while} loop in Algorithms \ref{alg1} and \ref{alg2} so that a single iteration is run
per time interval. Fig. \ref{fig:jstsp_fig_sg_ista_fista_admm_online}
compares the resulting MSE curves as the error evolves with time, showing that
the inexact online FISTA algorithm achieves the 
best error performance. The MSE performance degradation 
of Algorithm \ref{alg3} relative to its (exact) counterpart Algorithm \ref{alg2}
is depicted in Fig. \ref{fig:jstsp_fig_fista_iterations}, as a function of the number of inner iterations
$k$.


%

\begin{figure}[t]
\centering
\includegraphics[scale=0.7]{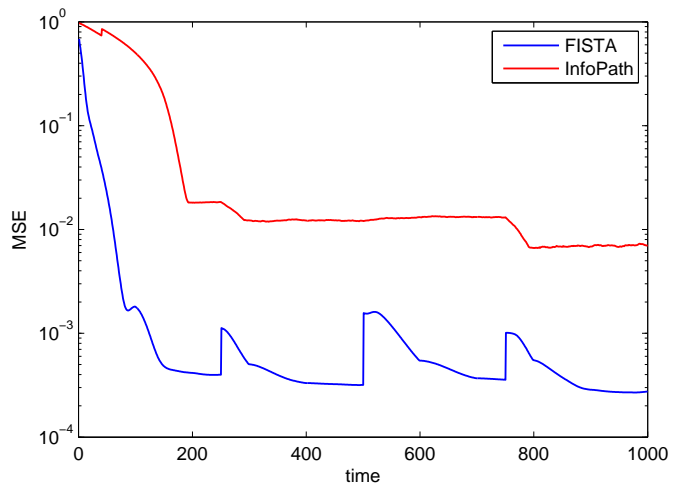}
\caption{MSE performance evolution of the pseudo real-time FISTA (Algorithm \ref{alg2}) 
compared with the InfoPath algorithm in~\cite{gomez}.}
\label{fig:jstsp_fig_fista_vs_infopath}
\end{figure}

\noindent\textbf{Comparison with~\cite{gomez}.} 
The proposed Algorithm \ref{alg2} is compared here to the method of~\cite{gomez}, 
which  does not explicitly account for external influences and edge sparsity. To this end, the stochastic-gradient descent 
algorithm (a.k.a. ``InfoPath") developed in~\cite{gomez} is run using the generated synthetic
data with non-smooth edge variations. Postulating an exponential transmission model,
the dynamic network is tracked by InfoPath by performing MLE 
of the edge transmission rates (see~\cite{gomez} for
details of the model and the algorithm). Note that the postulated model
therein differs from \eqref{eq:mps3}, used here to generate the network data.
Fig. \ref{fig:jstsp_fig_fista_vs_infopath} depicts the MSE
performance of ``InfoPath" compared against FISTA.
Apparently, there is an order of magnitude reduction in MSE
by explicitly modeling external sources of influence and leveraging the attribute
of sparsity.

\subsection{Real data}
\label{ssec:real}

\noindent\textbf{Dataset description.} The real data used was
collected during a prior study by monitoring blog posts and news articles 
for memes (popular textual phrases) appearing within a set
of over $3.3$ million websites~\cite{gomez}. Traces of information cascades
were recorded over a period of one year, from March $2011$
till February $2012$; the data is publicly available from~\cite{data}. The time when each website mentioned
a specific news item was recorded as a Unix timestamp
in hours (i.e., the number of hours since midnight on January $1$, $1970$).
Specific globally-popular topics during
this period were identified and cascade data for the top $5,000$
websites that mentioned memes associated with them were retained.
The real-data tests that follow focus on the topic ``Kim Jong-un", the 
current leader of North Korea whose popularity rose after the death of 
his father and predecessor, during the observation period.

Data was first pre-processed and filtered so that only (significant) 
cascades that propagated to at least $7$
websites were retained. This reduced the dataset significantly
to the $360$ most relevant websites over which
$466$ cascades related to ``Kim Jong-un" propagated. 
The observation period was then split into
$T = 45$ weeks, and each time interval was set to one week. In addition,
the observation time-scale was adjusted to start at the beginning
of the earliest cascade.

Matrix $\Y^t$ was constructed by setting $y_{ic}^t$
to the time when website $i$ mentioned phrase $c$ if
this occurred during the span of week $t$. Otherwise
$y_{ic}^t$ was set to a large number, $100 t_{\text{max}}$,
where $t_{\text{max}}$ denotes the largest timestamp in
the dataset. Typically the entries of matrix $\X$ 
capture prior knowledge about the susceptibility
of each node to each contagion. For instance,
the entry $\x_{ic}$ could denote the online
search rank of website $i$ for a search keyword
associated with contagion $c$. In the absence of
such real data, the entries of $\X$ were generated
randomly from a uniform distribution over 
the interval $[0,0.01]$.

\begin{figure*}[t]
\begin{minipage}[b]{.33\textwidth}
  \centering
  \includegraphics[width=5.57cm]{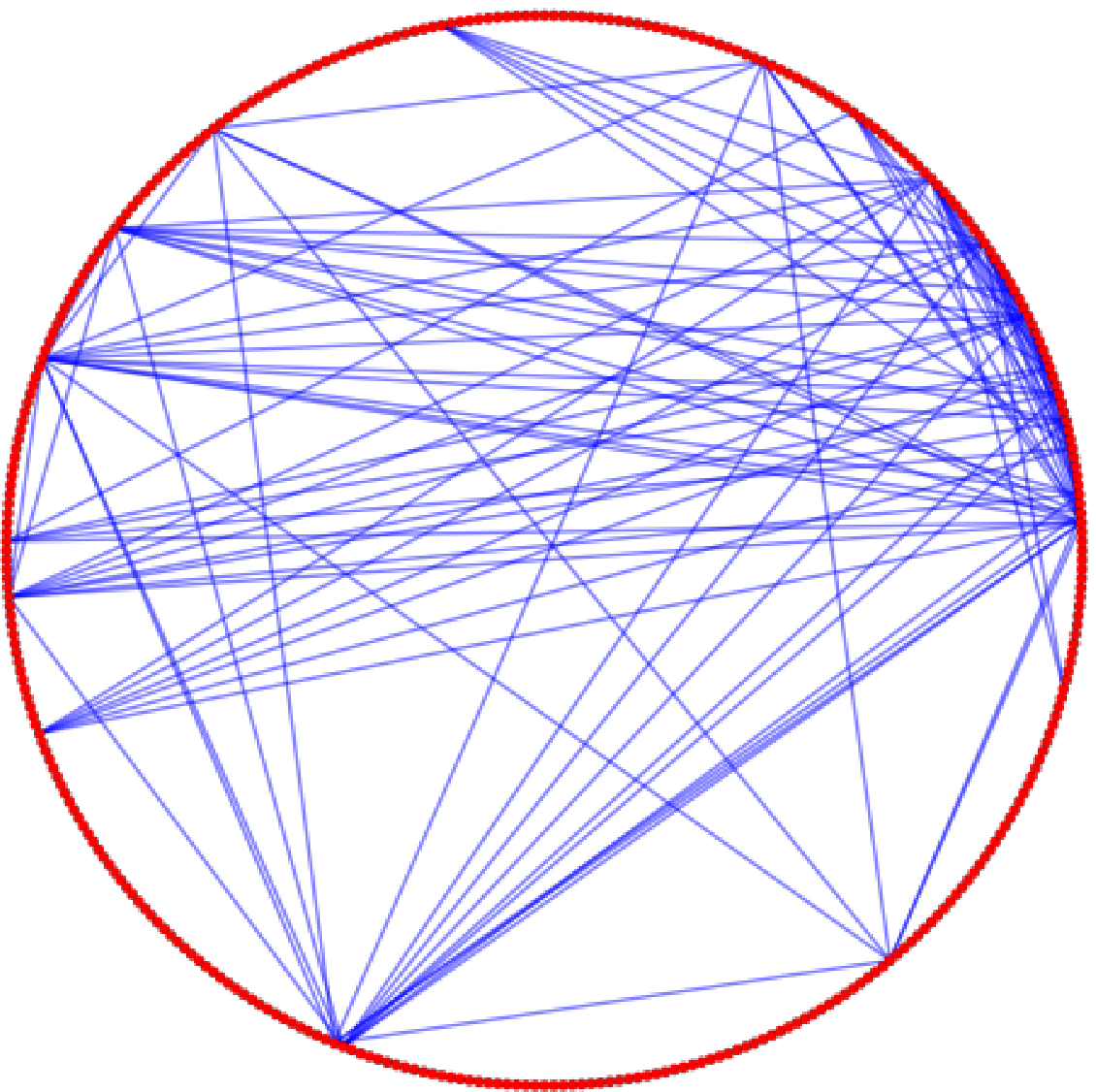}
  \centerline{(a) $ t = 10$} 
\end{minipage}
%
\begin{minipage}[b]{.33\textwidth}
  \centering
  \includegraphics[width=5.8cm]{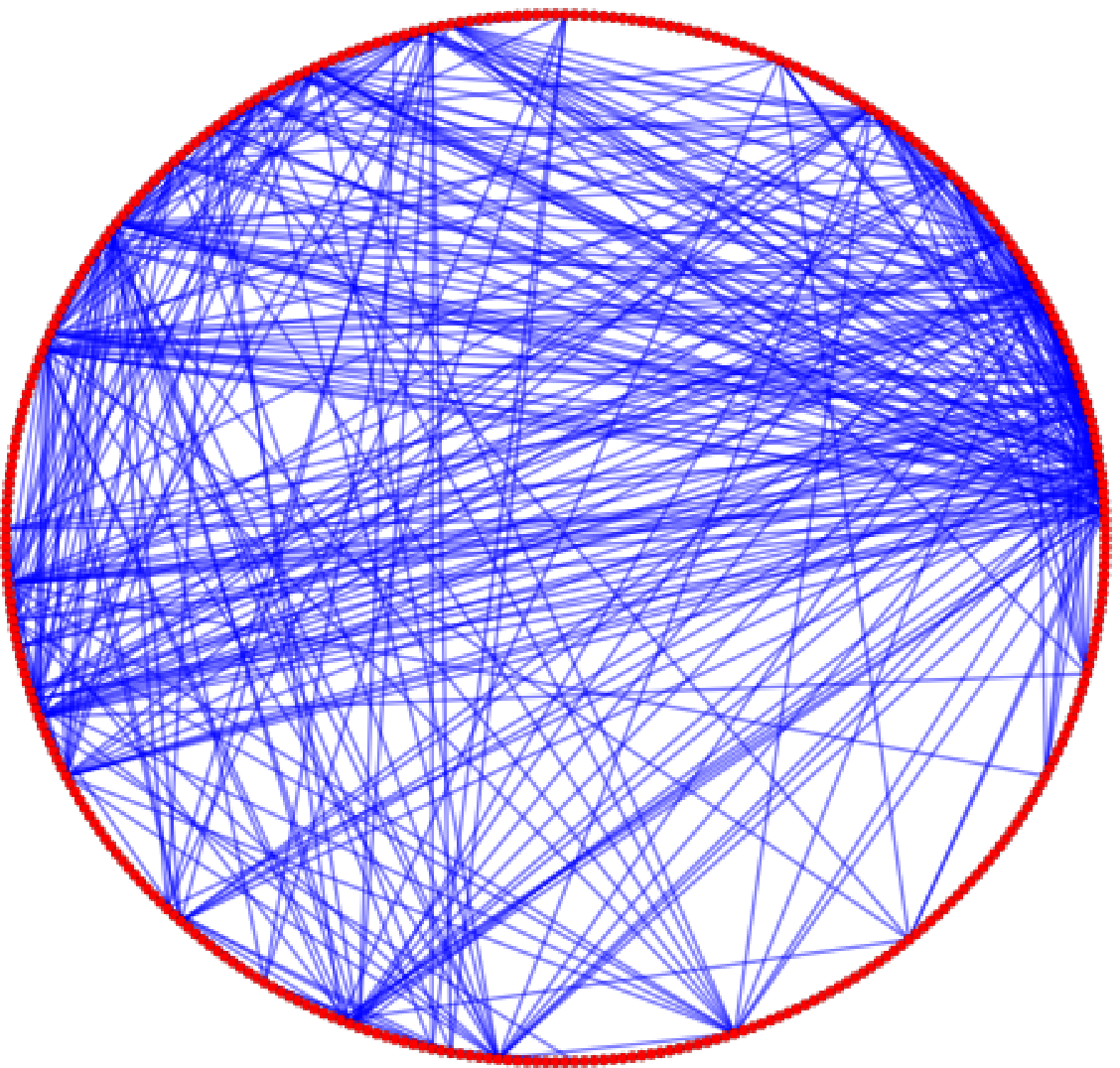}
  \centerline{(b) $t = 30$} 
\end{minipage}
%
\begin{minipage}[b]{0.33\textwidth}
  \centering
 \includegraphics[width=5.8cm]{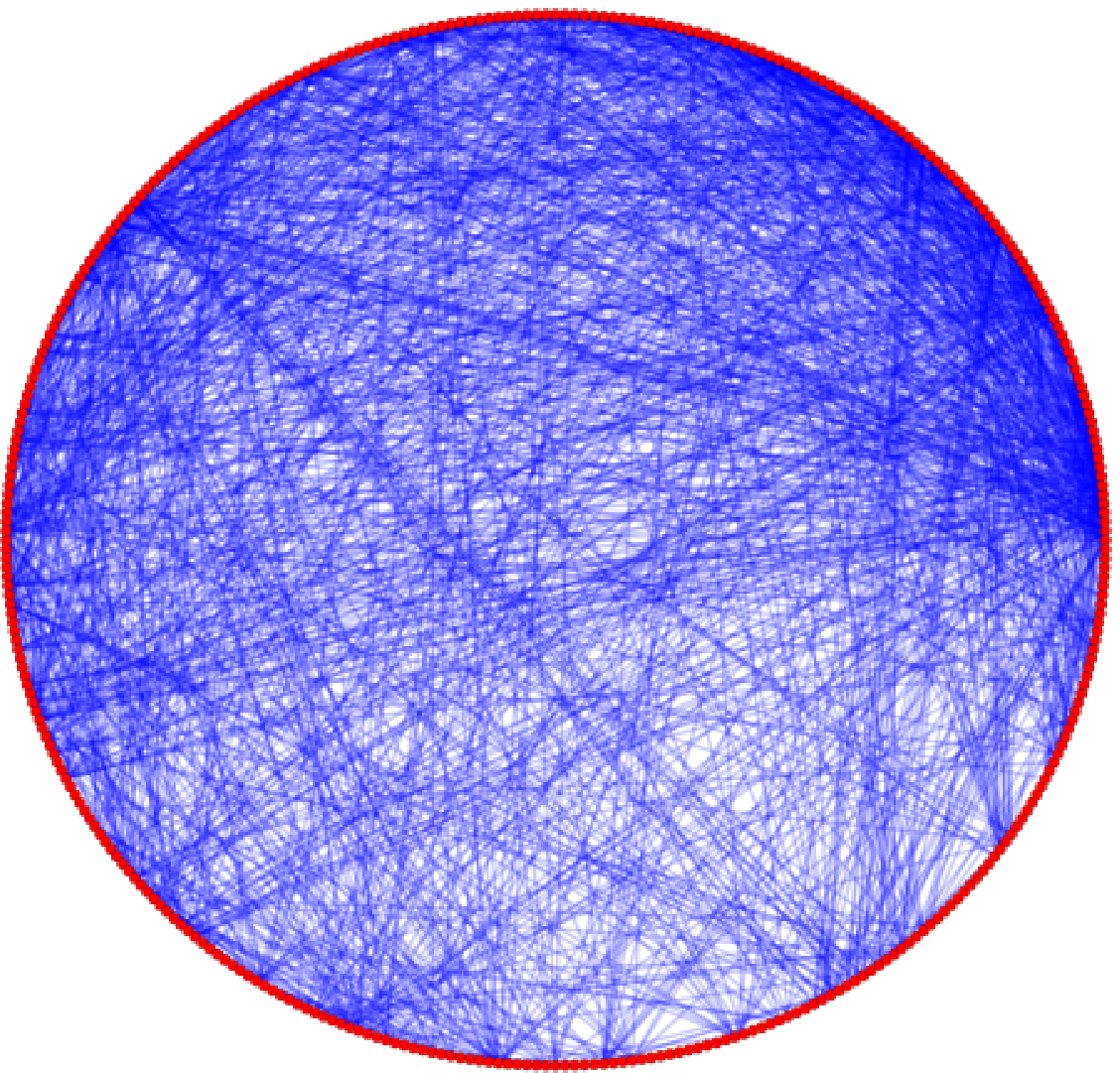}
  \centerline{(c) $t = 40$} 
\end{minipage}

\caption{Visualization of the estimated networks obtained by tracking those 
information cascades related to the topic ``Kim Jong-un". The abrupt increase
in network connectivity can be explained by three key events:  i) Kim Jong-un was appointed as the vice chairman
of the North Korean military commission ($t=28$); ii) Kim Jong-il died ($t=38$); 
and iii) Kim Jong-un became the ruler of North Korea ($t=40$).}
\label{fig:kimjongun}

\end{figure*}

\noindent\textbf{Experimental results.} Algorithm \ref{alg2}
was run on real data with $\beta = 0.9$ and $\lambda_t = 100$.
Fig. \ref{fig:kimjongun} depicts circular drawings of the 
inferred network at $t = 10$, $t=30$, and  $t = 40$ weeks. 
Little was known about Kim Jong-un during the first $10$ weeks 
of the observation period. However, speculation about the possible
successor of the dying North Korean ruler, Kim Jong-il, rose 
until his death on December 17, 2011 (week $38$).
He was succeeded by Kim Jong-un on December 30, 2011 (week $40$). 
The network visualizations show an increasing number
of edges over the $45$ weeks, illustrating the growing interest of
international news websites and blogs in the new ruler. Unfortunately,
the observation horizon does not go beyond $T=45$ weeks. A longer span of
data would have been useful to investigate at what rate did the global
news coverage on the topic eventually subside.

\begin{figure}[t]
\centering
\includegraphics[scale=0.55]{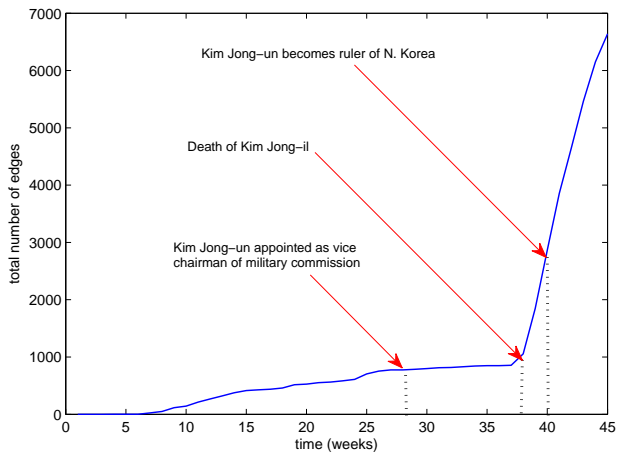}
\caption{Plot of total number of inferred edges per week.}
\label{fig:jstsp_fig_kim_edges_vs_time}
\end{figure}

Fig. \ref{fig:jstsp_fig_kim_edges_vs_time}
depicts the time evolution of the  total number of edges in the inferred dynamic network. 
Of particular interest are the weeks 
during which: i) Kim Jong-un was appointed as the vice chairman
of the North Korean military commission; ii) Kim Jong-il died; 
and iii) Kim Jong-un became the ruler of North Korea. These 
events were the topics of many online news articles
and political blogs, an observation that is reinforced by 
the experimental results shown in the plot.

\section{Concluding Summary}
\label{sec:conclusion}
A dynamic SEM was proposed in this paper for network topology inference, 
using timestamp data for propagation of contagions typically observed
in social networks. The model explicitly captures
both topological influences and external sources of information diffusion over the unknown
network. Exploiting the inherent edge sparsity typical of 
large networks, a computationally-efficient proximal gradient algorithm with
well-appreciated convergence properties was developed
to minimize a suitable sparsity-regularized exponentially-weighted LS estimator. 
Algorithmic enhancements were proposed, that pertain to accelerating convergence and  
performing the network topology inference task in real time. In addition, 
reduced-complexity stochastic-gradient iterations were outlined and showed to attain
worthwhile performance.

A number of experiments conducted on synthetically-generated 
data demonstrated the effectiveness of the proposed 
algorithms in tracking dynamic and sparse networks. 
Comparisons with the InfoPath
algorithm revealed a marked improvement
in MSE performance attributed to the explicit modeling of external influences
and leveraging edge sparsity. Experimental results on a real dataset focused
on the current ruler of North Korea successfully showed a sharp increase
in the number of edges between media websites in agreement
with the increased media frenzy following his 
ascent to power in 2011. 

The present work opens up multiple directions for exciting follow-up work. 
Future and ongoing research includes: 
i) investigating the conditions for identifiability of sparse and 
dynamic SEMs, as well as their statistical consistency properties tied to
the selection of $\lambda_t$; ii)
formally establishing the convergence of the (inexact) real-time algorithms
in a stationary network setting, and tracking their MSE performance under
simple models capturing the network variation; iii) devising
algorithms for MLE of dynamic SEMs and comparing the performance of the LS
alternative of this paper; iii) generalizing the SEM using kernels
or suitable graph similarity measures to enable network topology forecasting;
and iv) exploiting the parallel structure of the algorithms to devise MapReduce/Hadoop 
implementations scalable to million-node graphs.


\bibliographystyle{IEEEtranS}
\bibliography{IEEEabrv,biblio}

\end{document}